\documentclass[10pt,prl,aps,twocolumn,showpacs,floatfix]{revtex4}

\usepackage{graphicx}
\usepackage{amssymb}
\usepackage{subfigure}
\usepackage{color}
\usepackage{amsmath} 
\begin{document}

\title{Universal N\'eel Temperature in Three-Dimensional Quantum Antiferromagnets}

\author{Songbo Jin and Anders W. Sandvik}
\affiliation{Department of Physics, Boston University, 590 Commonwealth Avenue, Boston, Massachusetts 02215, USA}

\begin{abstract}
We study three-dimensional dimerized $S=1/2$ Heisenberg antiferromagnets, using quantum Monte Carlo simulations of systems with three 
different dimerization patterns. We propose a way to relate the N\'eel temperature $T_N$ to the staggered moment $m_s$ of the ground state. 
Mean-field arguments suggest $T_N \propto m_s$ close to a quantum-critical point. We find an almost perfect universality (including 
the prefactor) if $T_N$ is normalized by a proper lattice-scale energy. We show that the temperature $T^*$ at which the magnetic susceptibility 
has a maximum is a good choise, i.e., $T_N/T^*$ versus $m_s$ is a universal function (also beyond the linear regime). These 
results are useful for analyzing experiments on systems where the spin couplings are not known precisely, e.g., $\mathrm{TlCuCl_3}$.
\end{abstract}

\date{\today}


\maketitle
Quantum fluctuation can drive continuous phase transitions between different kinds of ground states of many-body systems. While transitions 
taking place at temperature $T>0$ are controlled by thermal fluctuations, quantum fluctuations also play a role here. Quantum-critical scaling 
can often be observed throughout a wide region (the quantum-critical ``fan'') extending out from the quantum-critical point $(g_c,T=0)$ into 
the plane $(g,T>0)$ \cite{Chakravarty89,Chubukov94,Shevchenko00,Sachdevbook},  where $g$ is the parameter tuning the strength of the quantum 
fluctuations. In addition, the quantum fluctuations of course also strongly affect the critical temperature $T_c$, because $T_c \to 0$ as $g \to g_c$. 
One can regard the quantum fluctuations as reducing the order at low temperature ($T \ll T_c$), with the thermal fluctuations eventually 
destroying it as $T \to T_c$, but precisely how the two kinds of fluctuations act in conjunction with each other to govern $T_c$ is not known 
in general.

\begin{figure}
\center{\includegraphics[width=8cm, clip]{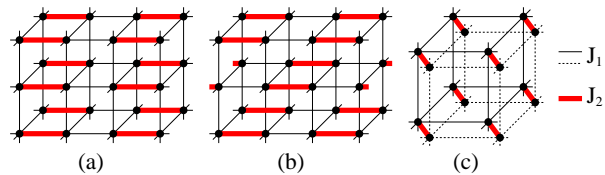}}
\vskip-2mm
\caption{(Color online) Dimerized 3D lattices; (a) columnar dimers, (b) staggered dimers, and (c) double cube. For a system of length $L$, 
the number of spins is $N=L^3$ in (a) and (b), and $N=2L^3$ in (c). The two different coupling strengths $J_1$ and $J_2$ are indicated by 
thin (black dashed and solid) and thick (red) lines, respectively.}
\label{fig1}
\vskip-4mm
\end{figure}

We will here discuss manifestations of the interplay of quantum and thermal fluctuations for $0 < T<T_N$ in three-dimensional (3D) $S=1/2$ quantum 
antiferromagnets with Heisenberg interactions. In these systems one can vary the critical N\'eel-ordering temperature, $T_N$, and ultimately achieve a quantum 
phase transition ($T_N \to 0$), by considering dimerized couplings, such that each spin belongs exactly to one dimer and the intra- and inter-dimer 
couplings are different. The Hamiltonian for such models can be generically written as
\begin{equation}
H=J_1\sum_{\langle i,j\rangle_1}\mathbf{S}_{i}\cdot\mathbf{S}_{j} + J_2 \sum_{\langle i,j\rangle_2}\mathbf{S}_{i}\cdot\mathbf{S}_{j},
\label{eq:ham}
\end{equation} 
where $\langle i,j\rangle_a$ denotes a pair of spins coupled at strength $J_a$, with $a=1$ and $a=2$ corresponding to inter- and intra-dimer 
bonds, respectively. Three examples of such dimerized 3D lattices are shown in Fig.~\ref{fig1}. In (a) and (b) the spins form simple cubic lattices, 
and each nearest-neighbor site pair is coupled either by $J_1$ or $J_2$. In (c) two different cubes each have all $J_1$ couplings, and pairs of spins 
in different cubes form the $J_2$-coupled dimers. We will use the ratio $g=J_2/J_1$ as the tuning parameter. When $g\approx 1$ the system is N\'eel-ordered 
at $T=0$ and when $g\to \infty$ it decouples to form a set of independent dimers, with the ground state becoming a trivial quantum paramagnet with a 
singlet-product ground state. The system for any $J_1>0$ and $J_2>0$ is accessible to unbiased numerical studies with efficient quantum Monte Carlo 
(QMC) methods with loop updates \cite{Sandvik99,Evertz03,Sandvik10}.

Analogous dimerized Heisenberg models have been studied extensively with QMC in two dimensions, where there is order only at $T=0$ (for $g<g_c$) 
and the nature of the quantum-critical point and its associated scaling fan has been the main focus of interest \cite{Sandvik10,layers}. Some 
simulations have also been previously carried out for 3D dimerized models \cite{Troyer97,Nohadani05,Yao07}. Here we report calculations uncovering 
universal aspects of the ordering temperature, from systems close to the quantum-critical point to deep inside the N\'eel phase. We develop a scaling 
procedure of direct relevance to experiments. Our results also provide new insights into the relevant energy scales present in the 3D N\'eel state 
and demonstrate an effective decoupling of thermal and quantum fluctuations.

{\it Experimental issues.}---The best experimental realization so far of a dimerized system with a quantum phase transition is $\mathrm{TlCuCl_3}$ under 
pressure \cite{Cavadini01,Ruegg04,Ruegg08}. The spin dimers here form on pairs of Cu atoms that can clearly be identified as the most strongly coupled 
neighbors. The inter-dimer couplings are, however, more complicated than in the simple nearest-neighbor Hamiltonian (\ref{eq:ham}). There are several significant 
exchange constants but their exact values are not known (although they have been estimated based on approximate calculations of the magnon
dispersion, which can be compared with experiments  \cite{Cavadini01,Matsumoto04}). The dimers nevertheless form a 3D network, and one can expect the same ground 
state phases and phase transitions as with the  simplified Hamiltonian (\ref{eq:ham}). Under ambient pressure, $\mathrm{TlCuCl_3}$ exhibits no magnetic 
order, but beyond a critical pressure antiferromagnetic order emerges continuously. The interpretation of this is that one or several of the inter-dimer 
couplings become strong enough for N\'eel order to form. The observed longitudinal and transversal excitation energies agree well with predictions based 
on $O(3)$ symmetry breaking and Goldstone modes \cite{Matsumoto04,Sachdev09}.

The fact that the microscopic spin-spin couplings in $\mathrm{TlCuCl_3}$, and how they depend on pressure, are not known accurately is a complication 
when comparing experimental results with calculations for a specific model Hamiltonian. In this situation it is useful to make comparisons that do 
not require any explicit knowledge of the couplings. Here we will investigate how the N\'eel temperature is related to the staggered magnetization 
$m_s$ at $T=0$. Based on unbiased QMC calculations for the three different dimerized models defined in Eq.~(\ref{eq:ham}) 
and Fig.~\ref{fig1}, we show that the curve $T_N(m_s)$ exhibits a remarkable universality when properly normalized, not just close to the 
quantum-critical point but extending to strongly ordered systems. Our results give a parameter-free scaling function that can be compared
with experiments.

{\it Quantum Monte Carlo calculations.}---We have used the stochastic series expansion (SSE) QMC method with very efficient loop updates
\cite{Sandvik99,Evertz03,Sandvik10} to calculate the squares $\langle m_z^2\rangle$ and $\langle m_{sz}^2\rangle$ of the $z$-components of the uniform 
and staggered magnetizations, 
\begin{equation}
m_z=\frac{1}{N}\sum_{i=1}^{N}S^z_i,~~~~~~m_{sz}=\frac{1}{N}\sum_{i=1}^{N}\phi_iS^z_i,
\label{eq:ms}
\end{equation}
where the phases $\phi_i=\pm 1$ correspond to the sublattices of the bipartite systems in Fig.~\ref{fig1}. The uniform susceptibility is 
$\chi=\langle m_z^2 \rangle/(TN)$. We also study the Binder ratio, $R_2=\langle m_{sz}^4 \rangle/\langle m_{sz}^2 \rangle ^2$, and the 
spin stiffness constants $\rho_s^\alpha$ in all lattice directions ($\alpha=x,y,z$), $\rho_s^\alpha=d^2 E(\theta_\alpha)/d \theta_\alpha^2$, 
where $E$ is the internal energy per spin and $\theta_\alpha$ a uniform twist angle imposed between spins in planes perpendicular to the $\alpha$ 
axis. The stiffness constants can be related to winding number fluctuations in the simulations \cite{Sandvik10}. 

\begin{figure}
\center{\includegraphics[width=6.5cm, clip]{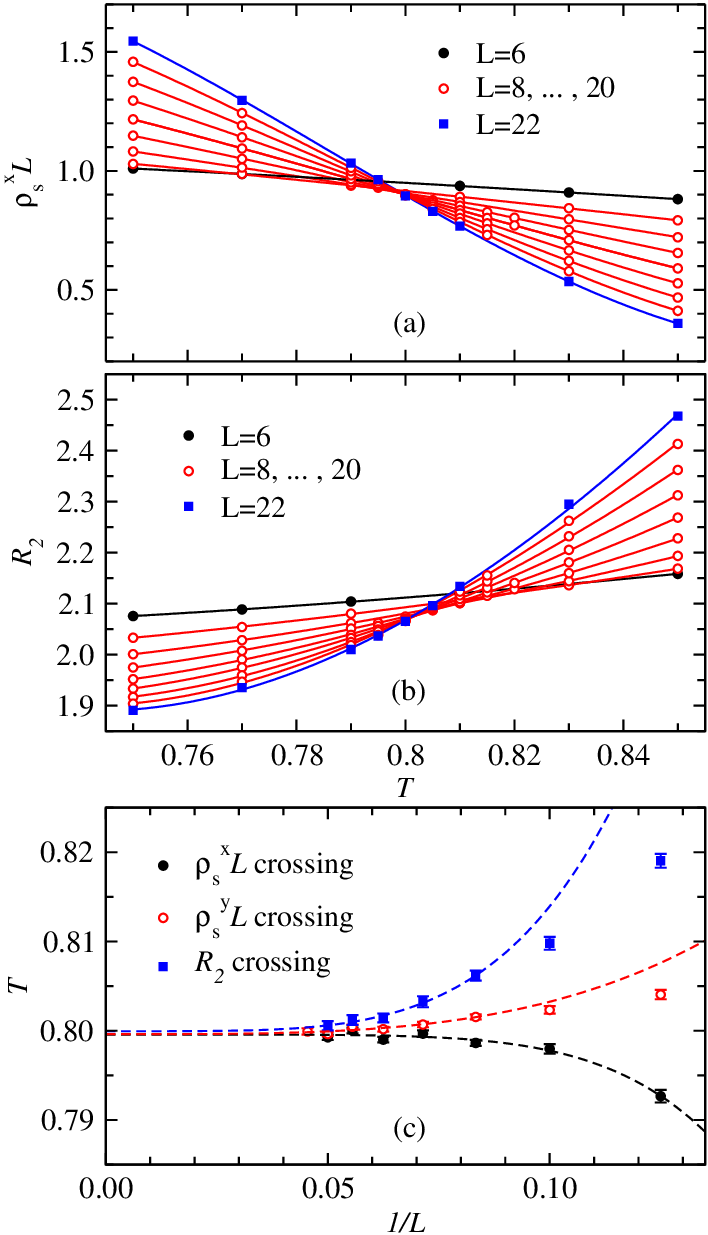}}
\vskip-1mm
\caption{(Color online) Procedures used to extract the critical temperature $T_N$. (a) and (b) show $\rho_s^xL$ and $R_2$, respectively,
for the columnar dimer model at coupling ratio $g=3.444$. The error bars are smaller than the symbols. Using polynomial fits to data for two 
lattice sizes, $L$ and $L+2$, crossing points between the curves are extracted. Results are shown in (c), along with fits of the form $T_N(L)=T_N(\infty)+a/L^w$ 
(to the large-$L$ data for which this form obtains). Extrapolations of the three quantities give $T_N=0.7996(3)$, $0.7996(6)$, and $0.7999(5)$
for $L \to \infty$, all consistent with each other within errors bars.}
\label{fig2}
\vskip-3mm
\end{figure}

We use standard finite-size scaling \cite{Sandvik10} to extract $T_N$. At $T_N$, the stiffness constants scale with the system length as 
$\rho_s^\alpha \propto L^{2-d}$, where the dimensionality $d=3$. Thus, $\rho_s^\alpha L$ should be size-independent at $T_N$, while this quantity 
vanishes (diverges) for $T>T_N$ ($T<T_N$). In practice, this means that curves versus $T$ (at fixed $g$) for two different system sizes $L$ cross 
each other at a point which drifts (due to scaling corrections) toward $T_N$ with increasing $L$. 
The dimensionless Binder ratio also has this kind of behavior and provides us with a different $T_N$
estimate to check for consistency. Figs.~\ref{fig2}(a,b) show examples of these crossing behaviors for $\rho_s^xL$ and $R_2$. The crossing points  
drift in different directions and bracket $T_N$. Fig.~\ref{fig2}(c) shows the $L$ dependence of crossing points extracted from data for $(L,L+2)$ 
system pairs, for $R_2$ and two different stiffness constants. Power-law fits are used to extrapolate to infinite size. The mutual
consistency of the $T_N$ value so obtained using different quantities gives us confidence in the accuracy of this procedure.

\begin{figure}
\center{\includegraphics[width=7.0cm,clip]{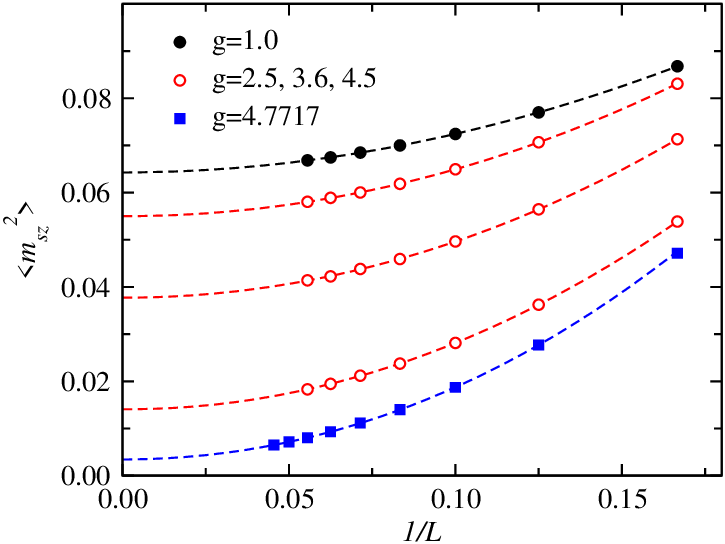}}
\vskip-1mm
\caption{(Color online) Extrapolation of the sublattice magnetization obtained in simulations with $T=J_1/L$ of the double-cube Heisenberg model 
at different coupling ratios $g$. The error bars are much smaller than the symbols. The fitting function used for $L\to \infty$  extrapolations
is $a+b/L^2+c/L^3$ (where we exclude the linear term because it comes out very close to zero in fits including it).}
\label{fig3}
\vskip-3mm
\end{figure}

To extract the $T=0$ sublattice magnetization, we carry out simulations at temperature $T=J_1/L$. Note that, in a N\'eel phase with $T_N>0$, any $T(L)$
such that $T(L\to \infty) \to 0$ can be used for extrapolations to the thermodynamic limit and $T=0$. Our choice is a natural way to to scale the temperature
since the lowest spin waves have energy $\propto 1/L$. We also did some calculations with $T=1/2L$ and obtained consistent extrapolated results. 
Examples of the $L$ dependence are shown in Fig.~\ref{fig3} for the double-cube model at several different coupling 
ratios. Taking into account rotational averaging in spin space, the final result for the sublattice magnetization is given by the $L\to \infty$ 
extrapolated $\langle m_{sz}^2\rangle$ (for which we use a polynomial fit, as shown in Fig.~\ref{fig3}); $m_s=\sqrt{3 \langle m_{sz}^2\rangle}$.

\begin{figure}
\center{\includegraphics[width=6.5cm, clip]{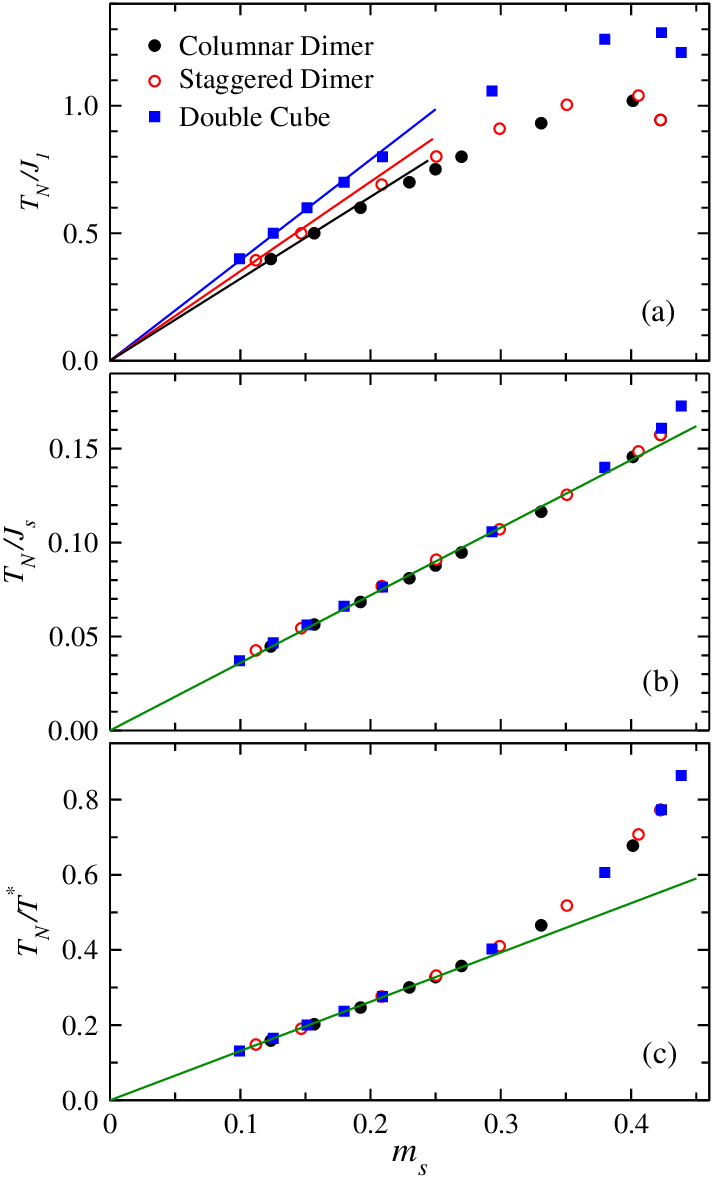}}
\vskip-1mm
\caption{(Color online) The N\'eel temperature $T_N$ versus the sublattice magnetization for the three different dimerized models and with $T_N$
normalized in three different ways. $T_N$ is measured in units of (a) the inter-dimer coupling $J_1$, (b) the total coupling $J_s$ per spin, 
(c) the peak temperature $T^*$ of the susceptibility. A linear dependence obtains in all cases for small to moderate $m_s$, as indicate 
by fitted lines. Note that $m_s \le 1/2$ for $S=1/2$.}
\label{fig4}
\vskip-3mm
\end{figure}

{\it Universality of $T_N$ versus $m_s$.}---Following the above procedures, we have calculated $T_N$ and $m_s$ accurately for all three dimer 
models at several coupling ratios $g$, from close to $g_c$ to deep inside the N\'eel phase. We graph $T_N$ versus $m_s$ in Fig.~\ref{fig4}. 
$T_N$ is scaled by three different energy units; the inter-dimer coupling 
$J_1$ in (a), the sum of couplings $J_s$ connected to each spin in (b), and the temperature $T^*$ at which the susceptibility exhibits a peak in 
(c). Before discussing these normalizations of $T_N$ in detail, let us examine the reason for the linear behavior, $T_N \propto m_s$, seen in the 
QMC results for small [and in (b),(c) even quite large] $m_s$.

A semi-classical mean-field argument (inspired by the ``renormalized classical'' picture developed in two dimensions \cite{Chakravarty89}) leading 
to $T_N \propto m_s$ is the following: To compute $T_N$ in a classical system of spins of length $S$, one replaces the coupling of a spin ${\bf S}_0$ 
to the total spin of its neighbors $\delta$, $J\sum_\delta {\bf S}_\delta$, by the thermal average $J\sum_\delta \langle {\bf S}_\delta \rangle$. In 
the presence of quantum fluctuations, this mean field seen by ${\bf S}_0$ is reduced, which is taken into account by a renormalization; 
$\langle {\bf S}_\delta\rangle \to (m_s/S)\langle {\bf S}_\delta\rangle$. The thermal fluctuations are, thus, added on top of the quantum fluctuations 
at $T=0$, under the assumption that the quantum effects will not change appreciably for $T>0$ (i.e., the thermal fluctuations are 
assumed to be solely responsible for further reducing the order). Note that ${\bf S}_0$ should not be renormalized here, but is computed as a 
thermal expectation value and should satisfy the self-consistency condition $\langle {\bf S}_\delta\rangle=\langle {\bf S}_0 \rangle$. The final 
magnetization curve is given by $(m_s/S)\langle {\bf S}_0\rangle$. In this procedure of decoupling the classical and quantum fluctuations, one 
clearly effectively has $J \to (m_s/S)J$ and, thus, $T_N \propto m_s$. 

The assumption that the quantum renormalization factor $m_s/S$ is $T$-independent up to $T_N$ can be valid only if $T_N$ is small. The energy scale 
in which to measure $T_N$ when stating this condition should be dictated by the spin-wave velocity, which stays non-zero at the quantum-critical point 
\cite{Kulik11} [i.e., not by the long-distance energy scale $\rho_s(T=0)$, which vanishes as $g \to g_c$ and is unrelated to the density of thermally 
excited spin waves]. A linear dependence is seen in Fig.~\ref{fig4} up to rather large values of $m_s$ (where $T_N \sim J_1$). A linear dependence 
was also recently found in the columnar dimer model based on high-$T$ expansions \cite{Oitmaa11} (with much larger error bars).

Returning now to the issue of how to best normalize $T_N$, we note that in Fig.~\ref{fig4}(a), where the inter-dimer coupling $J_1$ is
used, the curve for the double-cube model is significantly above the other two. This is clearly because the constant $J_1$ does not account for
the different average couplings in the models. Using instead the sum $J_s$ of couplings connected to each spin, i.e., $J_s=5+g$ for the columnar
and staggered dimers and $6+g$ for the double cube (setting $J_1=1$), the curves, shown in Fig.~\ref{fig4}(b), collapse almost on top of 
each other. Note that also the curves for the columnar and staggered dimers are closer to each other than in Fig.~\ref{fig4}(a), although they 
have the same definition of $J_s$. This can be the case because $J_s$ rescales the curves non-uniformly, since $m_s(g)$ and, therefore, $J_s(m_s)$, 
is different for the two models. The linearity of $T_N/J_s$ versus $m_s$ is also much clearer than before and extends
all the way up to $m_s \approx 0.3$.

\begin{figure}
\center{\includegraphics[width=6.25cm, clip]{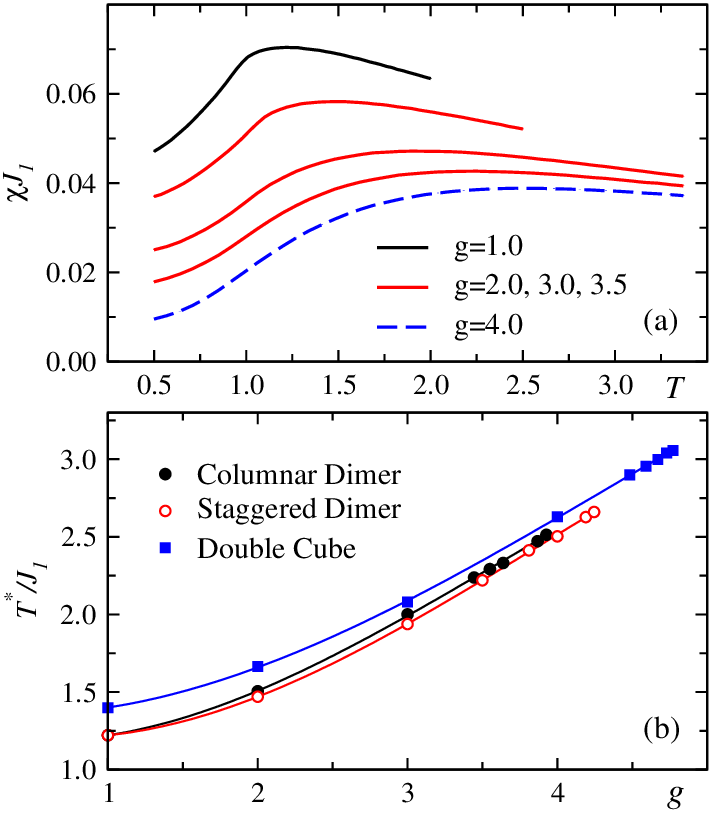}}
\vskip-2mm
\caption{(Color online) (a) Susceptibility versus temperature of the staggered dimer model at different coupling ratios. The system
size is $L=12$, for which the peak height and location are already $L\to \infty$ converged. (b) The peak temperature versus the coupling ratio
for the three different models.}
\label{fig5}
\vskip-3mm
\end{figure}

Although the data collapse is already quite good in $T_N/J_s$, we can do even better when normalizing with a physical quantity
that measures the effective lattice-scale energy. One such energy scale in antiferromagnets is the temperature at which the uniform magnetic 
susceptibility $\chi$ exhibits a peak. This peak is due to the cross-over from the high-$T$ Curie form to the low-$T$ weakly 
temperature dependent form typical of antiferromagnets. The peak temperature $T^*$, thus, reflects the short-distance energy scale at which 
antiferromagnetic correlations become significant. $T^*$ is often used experimentally to extract the value of the exchange constant, using, e.g., 
the ``Bonner-Fisher'' curve in one dimension \cite{Eggert96}. In spatially anisotropic systems such as the dimerized models we consider here, a 
natural assumption is that $T^*$ reflects an effective average coupling. In Fig.~\ref{fig5}(a) we show examples of the susceptibility close to its 
peak, and in (b) we show the dependence of $T^*$ on $g$ for all three models. Normalizing $T_N$ with $T^*$ leads to remarkably good data collapse, 
as shown in Fig.~\ref{fig4}(c). Deviations from a common curve are barely detectable. Although we cannot prove that this function is really universal 
for all 3D networks of dimers, the results are very suggestive of this. 

{\it Discussion.}---The universal behavior implies that the $T>0$ disordering mechanism in the 3D N\'eel state is completely governed by a single 
lattice-scale energy (which, as we have shown here, can be taken as the peak temperature $T^*$ of the susceptibility) and the $T=0$ sublattice 
magnetization $m_s$. The extended linear behavior seen in Figs.~\ref{fig4}(b,c) shows that the quantum and classical fluctuations at $T < T_N$ are completely 
decoupled all the way from $g=g_c$ (excluding $g_c$ itself, where $T_N=0$) to quite far away from the quantum-critical point. Depending on a lattice-scale energy 
instead of the quantum-critical spin stiffness, the linear behavior is not fundamentally a quantum-critical effect. We have discussed the linearity 
and decoupling of the fluctuations in terms of a semi-classical mean-field theory, the validity of which implies that the quantum-critical regime 
\cite{Chubukov94} commences only above $T_N$. Deviations from linearity at larger $m_s$ show that the quantum fluctuations are affected (become $T$-dependent) 
here, due to the high density of excited spin waves as $T \to T_N$ because $T_N$ is high. It is remarkable that this coupling of quantum and classical
fluctuations also takes place in an, apparently, universal fashion for different systems. It would be interesting to explain this more quantitatively, 
by deriving the full function $T_N$ versus $m_s$ analytically. Progress in the linear regime has been made recently in work parallel to ours \cite{Oitmaa11b}.

From a practical point of view, the data collapse of $T_N/T^*$ versus $m_s$ is very useful, because all the quantities involved can be measured 
experimentally and do not rely on microscopic details. The universal curve can be used to test the 3D Heisenberg scenario without adjustable 
parameters. The universality likely applies not only to dimer netwtorks, but also to systems where the quantum fluctuations are regulated in other ways. 

{\it Acknowledgments.}---We would like to thank Christian R\"uegg and Oleg Sushkov for stimulating
discussions. This work was supported by the NSF under Grant No.~DMR-1104708. 

\null

\end{document}